# Hierarchical Risk Parity and Minimum Variance Portfolio Design on NIFTY 50 Stocks

Jaydip Sen
*Dept of Data Science*
*Praxis Business School*
Kolkata, INDIA
email: jaydip.sen@acm.org

Saikat Mondal
*Dept of Data Science*
*Praxis Business School*
Kolkata, INDIA
email: sikatmo@gmail.com

Abhishek Dutta
*Dept of Data Science*
*Praxis Business School*
Kolkata, INDIA
email: lfcabhi@gmailcom

Sidra Mehtab
*Dept of Data Science*
*Praxis Business School*
Kolkata, INDIA
email: smehtab@acm.org

*Abstract*— Portfolio design and optimization have been always an area of research that has attracted a lot of attention from researchers from the finance domain. Designing an optimum portfolio is a complex task since it involves accurate forecasting of future stock returns and risks and making a suitable tradeoff between them. This paper proposes a systematic approach to designing portfolios using two algorithms, the critical line algorithm, and the hierarchical risk parity algorithm on eight sectors of the Indian stock market. While the portfolios are designed using the stock price data from Jan 1, 2016, to Dec 31, 2020, they are tested on the data from Jan 1, 2021, to Aug 26, 2021. The backtesting results of the portfolios indicate while the performance of the CLA algorithm is superior on the training data, the HRP algorithm has outperformed the CLA algorithm on the test data.

## I. Introduction

The design of optimized portfolios has remained a research topic of broad and intense interest among the researchers of quantitative and statistical finance for a long time. An optimum portfolio allocates the weights to a set of capital assets in a way that optimizes the return and risk of those assets. Markowitz in his seminal work proposed the mean-variance optimization approach which is based on the mean and covariance matrix of returns [1]. The algorithm, known as the *critical line algorithm* (CLA), despite the elegance in its theoretical framework, has some major limitations. One of the major problems being the adverse effects of the estimation errors in its expected returns and covariance matrix on the performance of the portfolio. Since it is extremely challenging to accurately estimate the expected returns of an asset from its historical prices, it is a popular practice to use either a minimum variance portfolio or an optimized risk portfolio with the maximum Sharpe ratio as better proxies for the expected returns. However, due to the inherent complexity, several factors have been used to explain the expected returns.

The *hierarchical risk parity* (HRP) algorithm attempts to address three major shortcomings of quadratic optimization methods which are particularly relevant to the CLA [2]. These problems are, instability, concentration, and under-performance. Unlike the CLA, the HRP algorithm does not require the covariance matrix of return values to be invertible. The HRP is capable of delivering good results even if the covariance matrix is ill-degenerated or singular, which is an impossibility for a quadratic optimizer. Interestingly, even though CLA's objective is to minimize the variance, the HRP is proven to have a lower probability of yielding lower out-of-sample variance than CLA. The major weakness of the CLA approach is that a small deviation in the forecasted future returns can make the CLA deliver widely different portfolios. Given the fact that future returns cannot be forecasted with sufficient accuracy, many researchers have proposed risk-based asset allocation using the covariance matrix of the returns [3]. However, this approach brings in another problem of instability. The instability arises because the quadratic programming methods require the inversion of a covariance matrix whose all eigenvalues must be positive. This inversion is prone to large errors when the covariance matrix is numerically ill-conditioned, i.e., when it has a high condition number [4]. The HRP is a new portfolio design method that addresses the pitfalls of the CLA using techniques of graph theory and machine learning [2]. This method uses the exploits the features of the covariance matrix without the requirement of its invertibility or positive-definiteness and effectively working on even a singular covariance matrix of returns.

This paper presents an algorithmic approach for building efficient portfolios by selecting stocks from seven sectors of the National Stock Exchange (NSE) of India. Based on the report of the NSE on July 30, 2021, the ten most significant stocks of six sectors and the 50 stocks included in the NIFTY 50 are first identified [5]. Portfolios are built using the CLA and the HRP algorithms for the seven sectors using the historical prices of the stocks from Jan 1, 2016, to Dec 31, 2020. The portfolios are backtested on the in-sample data of the stock prices from Jan 1, 2016, to Dec 31, 2020, and also on the out-of-sample data of stock prices from Jan 1, 2021, to Aug 26, 2021. Extensive results on the performance of the backtesting of the portfolios are analyzed to identify the better-performing algorithm for portfolio design.

The main contribution of the current work is threefold. First, it presents two different methods of designing robust portfolios, the CLA algorithm, and the HRP algorithm. These portfolio design approaches are applied to seven critical sectors of stocks of the NSE. The results may serve as a guide to investors in the stock market for making profitable investments in the stock market. Second, a backtesting method is proposed for evaluating the performances of the algorithms based on the daily returns yielded by the portfolios and their associated volatilities (i.e., risks). Since the backtesting is done both on the training and on the test data of the stock prices, the method is capable of identifying the more efficient algorithm both on the in-sample data and the out-of-sample data. Hence, a robust framework for evaluating different portfolios is demonstrated. Third, the returns of the portfolios on the seven sectors on the test data highlight the current profitability of investment and the volatilities of the sectors studied in this work. This information can be useful for investors.

The paper is organized as follows. In Section II, some existing works on portfolio design and stock price prediction are discussed briefly. Section III highlights the methodology followed. Section IV presents the results of the two portfolio design approaches on the seven sectors. Section VI concludes the paper.

## II. RELATED WORK

Due to the challenging nature of the problems and their impact on real-world applications, several propositions exist in the literature for stock price prediction and robust portfolio design for optimizing returns and risk in a portfolio. The use of predictive models based on learning algorithms and deep neural net architectures for stock price prediction is quite common [6-7]. Hybrid models are also demonstrated that combine learning-based systems with the sentiments in the unstructured non-numeric contents on the social web [8-9]. The use of multi-objective optimization, principal component analysis, and metaheuristics have also been proposed by some researchers in portfolio design [10-12]. Estimating volatility in future stock prices using GARCH has also been proposed in some work [13].

The current work presents two methods the CLA method and the HRP method to introduce robustness while maximizing the portfolio returns for eight sectors of the NSE of India. On the basis of the past prices of the stocks from Jan 2016 to Dec 2020, portfolios are designed using the CLA and the HRP algorithms for each sector. The backtesting of the portfolios is carried out on the in-sample data of stock prices from Jan 2016 to Dec 31, 2020, and also on the out-of-sample data from Jan 1, 2021, to Aug 26, 2021. The portfolios are evaluated on their volatilities and Sharpe Ratios on both training and the test data.

## III. METHODOLOGY

In this section, the seven-step approach adopted in designing the proposed system is discussed in detail. The seven steps are as follows.

**(1) *Choosing the sectors*:** Eight important sectors of NSE are first chosen. The selected sectors are: (i) auto, (ii) banking, (iii) FMCG, (iv) information technology (IT), (v) metal, (vi) pharmaceutical, (vii) realty, and (viii) NIFTY 50. NIFTY 50 contains the 50 most critical stock stocks from several sectors of the Indian stock market. For the other seven sectors, the top ten stocks are identified based on their contributions to the computation of the overall sector index to which they belong. The ten most significant stocks for each sector are chosen based on the report published by the NSE on July 30, 2021 [5].

**(2) *Data acquisition*:** The prices of the ten most critical stocks of the seven sectors and the 50 stocks listed in the NIFTY 50 are extracted using the *DataReader* function of the *data* sub-module of the *pandas_datareader* module in Python. The stock prices are extracted from Yahoo Finance, from Jan 1, 2016, to Aug 26, 2021. The stock price data from Jan 1, 2016, to Dec 31, 2020, are used for training the portfolio models, while the trained portfolios are tested on the data from Jan 1, 2021, to Aug 26. 2021. Although there are five features in the stock data, the variable *close* is chosen for analysis as the current work is based on univariate analysis.

**(3) *Derivation of the return and volatility*:** The changes in the *close* values for successive days in percentage represent the daily *return*. For computing the daily returns, the *pct_change* function of Python is used. Based on the daily returns, the daily and yearly volatilities of the stocks of every sector are computed. Assuming that there are 250 operational days in a calendar year, the annual volatility values for the stocks are arrived at by multiplying the daily volatilities by a square root of 250.

**(4) *Construction of the optimum risk portfolios*:** At this step, the least risk portfolios are designed first. The portfolio with the minimum variance is referred to as the minimum variance portfolio. The *efficient frontier* for a given sector represents the contour of a large number of portfolios on which the returns and the risks are plotted along the *y*-axis and the *x*-axis, respectively. The points on an efficient frontier have the property that they are the portfolios that yield the highest return for specified risk, or they introduce the lowest risk for a given return. As the risk is depicted on the horizontal axis, the left-most point on the efficient frontier is the minimum risk portfolio. For plotting the efficient frontier for a sector, weights are assigned randomly to its stocks over a loop iterating for 10,000 rounds in a Python program. However, the minimum risk portfolios are rarely adopted in practice, and a trade-off between the risk and return is done. For optimizing the risk, Sharpe Ratio (SR) is used. SR of a portfolio is the ratio of the difference between its return and that of the risk-free one, to its standard deviation [2]. It measures the marginal increase in return for a unit increase in risk. Essentially, SR optimizes the return and the risk by yielding a substantially higher return with a very marginal increase in the risk. The portfolio with a risk of 1% is assumed to be risk-free. The portfolio with the maximum Sharpe Ratio is the optimum-risk portfolio, identified by the *idmax* function in Python. This algorithm used in designing this portfolio is referred to as the *critical line algorithm* (CLA). The CLA portfolios for all nine sectors are built on Jan 1, 2021, based on the training data from Jan 1, 2016, to Dec 31, 2020. At the end of this step, the weights allocated by the CLA portfolio to all the stocks of the nine sectors are available.

**(6) *Hierarchical Risk Parity portfolio design*:** As an alternative to the CLA algorithm for portfolio design, the hierarchical risk parity (HRP) algorithm-based portfolios are designed for the eight sectors. The HRP algorithm works in three phases: (a) *tree clustering*, (b) *quasi-diagonalization*, and (c) *recursive bisection*. These steps are briefly described in the following.

*Tree Clustering:* The tree clustering used in the HRP algorithm is an agglomerative clustering algorithm. To design the agglomerative clustering algorithm, a hierarchy class is first created in Python. The hierarchy class contains a dendrogram method that received the value returned by a method called linkage defined in the same class. The linkage method received the dataset after pre-processing and transformation and computes the minimum distances between stocks based on their return values. There are several options for computing the distance. However, the ward distance is a good choice since it minimizes the variances in the distance between two clusters in the presence of high volatility in the stock return values. In this work, the ward distance has been used as a method to compute the distance between two clusters. The linkage method performs the clustering and returns a list of the clusters formed. The computation of linkages is followed by the visualization of the clusters through a dendrogram. In the dendrogram, the leaves represent the individual stocks, while the root depicts the cluster containing all the stocks. The distance between each cluster formed is represented along the y-axis, longer arms indicate less correlated clusters and vice versa.

*Quasi-Diagonalization:* In this step, the rows and the columns of the covariance matrix of the return values of the stocks are reorganized in such a way that the largest values lie along the diagonal. Without requiring a change in the basis of the covariance matrix, the quasi-diagonalization yields a very important property of the matrix – the assets (i.e., stocks) with similar return values are placed closer, while disparate assets are put at a far distance. The working principles of the algorithm are as follows. Since each row of the linkage matrix merges two branches into one, the clusters ($C_{N-1}$, 1) and ($C_{N-2}$, 2) are replaced with their constituents recursively, until there are no more clusters to merge. This recursive merging of clusters preserves the original order of the clusters [2]. The output of the algorithm is a sorted list of the original stocks (as they were before the clustering).

*Recursive Bisection:* The quasi-diagonalization step transforms the covariance matrix into a quasi-diagonal form. It is proven mathematically that allocation of weights to the assets in an inverse ratio to their variance is an optimal allocation for a quasi-diagonal matrix [15]. This allocation may be done in two different ways. In the bottom-up approach, the variance of a contiguous subset of stocks is computed as the variance of an inverse-variance allocation of the composite cluster. In the alternative top-down approach, the allocation among two adjacent subsets of stocks is done in inverse proportion to their aggregated variances. In the current implementation, the top-down approach is followed. A python function *computeIVP* computes the inverse-variance portfolio based on the computed variances of two clusters as its given input. The variance of a cluster is computed using another Python function called *clusterVar*. The output of the *clusterVar* function is used as the input to another Python function called *recBisect* which computes the final weights allocated to the individual stocks based on the recursive bisection algorithm.

The HRP algorithm performs the weight allocation to *n* assets in the base case in time $T(n) = O(\log_2 n)$, while its worst-case complexity is given by $T(n) = O(n)$. Moreover, unlike, the CLA which is an approximate algorithm, the HRP is an exact and deterministic algorithm. The HRP portfolios for all nine sectors are built on Jan 1, 2021, based on the training data from Jan 1, 2016, to Dec 31, 2020. At the end of this step, the weights allocated by the HRP to all the stocks in the nine sectors are available.

**(7) Backtesting the portfolios on the training and test data:** In the final step, the CLA and HRP portfolios for each sector are backtested over the training and the test data set. For backtesting, the daily returns values are computed for each sector for both portfolios. For the purpose of comparison, the Sharpe Ratios and the aggregate volatility values for each sector are also computed for both the training and the test data. Since the portfolio that yields a higher value of Sharpe Ratio for a given sector is assumed to have performed better for that sector. For each sector, the portfolios that perform better on the training and the test data are identified. The HRP portfolios are expected to produce better performance than the CLA portfolios []. The results of the nine sectors are extensively analyzed to verify whether this hypothesis holds good.

## IV. RESULTS

This section presents the detailed results and analysis of the portfolios. The eight sectors which are selected for study are (i) auto, (ii) FMCG, (iii) banking, (iv) IT, (v) metal, (vi) pharmaceutical, (vii) realty, and (viii) NIFTY50. The optimum and eigen portfolios and the LSTM predictive model are implemented in Python language with TensorFlow and Keras libraries and executed on the Google Colab platform. The model is trained and validated over 100 epochs.

*Auto sector:* The top ten stocks and their weights for computing the index of this thematic sector based on the NSE's report on Jul 30, 2021, are the following: Maruti Suzuki: 19.98, Mahindra & Mahindra: 15.33, Tata Motors: 11.36, Bajaj Auto: 10.75, Hero MotoCorp: 7.73, Eicher Motors: 7.60, Bharat Forge: 4.18, Balkrishna Ind.: 4.15, Ashok Leyland: 4.12, and MRF: 3.58 [2]. Fig. 1 depicts the dendrogram of the clusters of stocks for the auto sector. Since the y-axis plots the *ward linkage* values, the longer the length of the arms, the larger is the distance and less compact is the cluster. For example, the cluster containing the Bajaj Auto and Hero MotoCorp stocks is the most compact one, while the one containing Balkrishna Ind. and MRF is the least in compactness. It is evident that the tree clustering for the HRP has created four clusters for the auto sector for the allocation of weights. Fig 2 depicts the weight allocation done by the CLA and the HRP approach. It is clear that the HRP approach has achieved more diversification in the portfolio as the allocation in the CLA is much more skewed. Fig. 3 and 4 show the results of backtesting for the training and the test data. These graphs plot the daily returns along the y-axis. The summary of the backtesting results is presented in Table 1. While the CLA algorithm is found to have produced a higher Sharpe Ratio (SR) and lower volatility for the training data, the HRP has outperformed it on the test data producing a higher value of SR, albeit with slightly higher volatility.

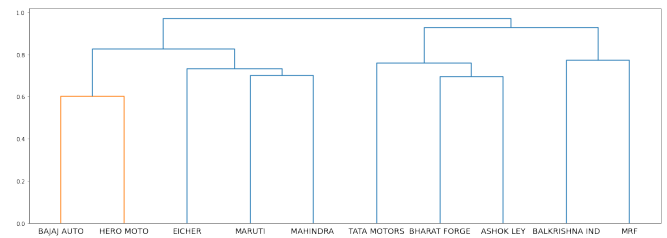

Fig. 1. The dendrogram of the agglomerative clustering of the auto sector stocks (Training data: Jan 1, 2016 – Dec 31, 2020)

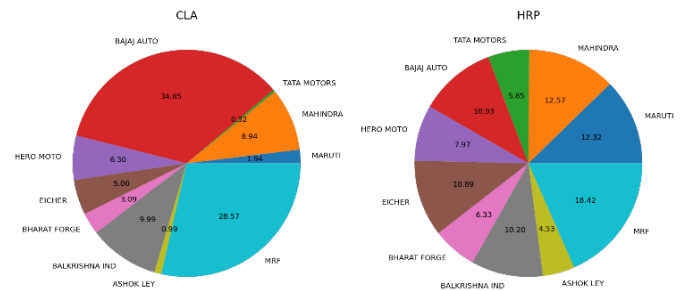

Fig. 2. Weight allocation to the auto sector stocks by the CLA and the HRP portfolio strategies (Training data: Jan 1, 2016 – Dec 31, 2020)

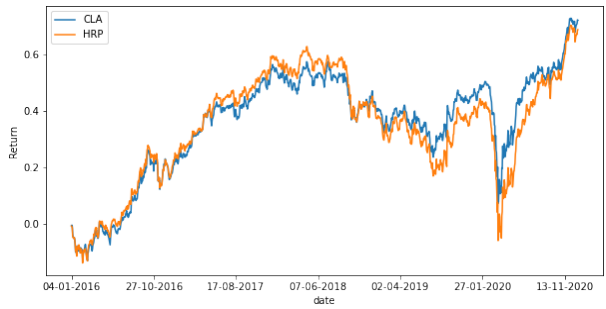

Fig. 3. Return yielded by the CLA and HRP portfolios on the training data for the auto sector stocks (Training data: Jan 1, 2016 – Dec 31, 2020)

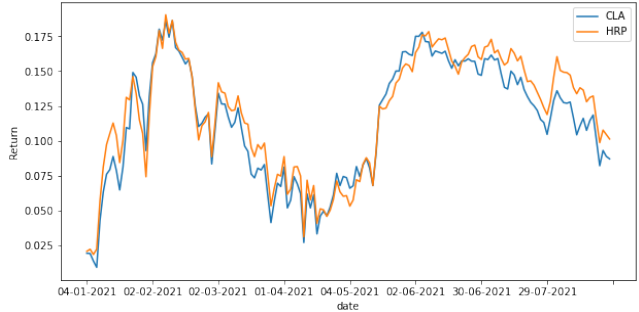

Fig. 4 Return from yielded by the CLA and HRP portfolios on the test data for the auto sector stocks (Test data: Jan 1, 2021 – Aug 26, 2021)

TABLE I. AUTO SECTOR PORTFOLIO PERFORMANCE

| Portfolio | Training Performance | | Test Performance | |
|---|---|---|---|---|
| | *Vol* | *Sharpe Ratio* | *Vol* | *Sharpe Ratio* |
| CLA | 0.2112 | 0.6977 | 0.2067 | 0.6641 |
| HRP | 0.2264 | 0.6210 | 0.2138 | 0.7468 |

*Banking sector:* The critical stocks of this sector are the following. HDFC Bank: 27.56, ICICI Bank: 22.91, State Bank of India: 12.42, Kotak Mahindra Bank: 11.77, Axis Bank: 11.49, IndusInd Bank: 5.20, AU Small Finance Bank: 2.35, Bandhan Bank: 1.73, Federal Bank: 1.53, and IDFC First Bank: 1.22 [2]. The results of the agglomerative tree clustering are depicted in Fig5. The Bandhan Bank stock could not participate in the clustering process due to the lack of availability of its historical prices. The first available data for its stock prices were March 27, 2018. Based on the threshold linkage value of the y-axis, the number of clusters will change. However, four clusters look very intuitive. Figs 6, 7, and 8 show the allocation of the weights by the two portfolios, and their performances on the training and on the test data, respectively. From Table II, it is observed that for the banking sector too, the HPR method has outperformed the CLA method on the test data by yielding a higher value for SR. However, the performance of the CLA is superior on the training data.

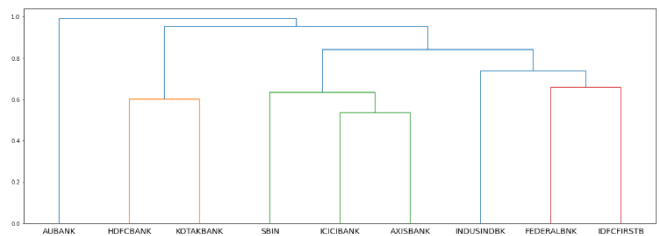

Fig. 5. The dendrogram of the agglomerative clustering of the banking sector stocks (Training data: Jan 1, 2016 – Dec 31, 2020)

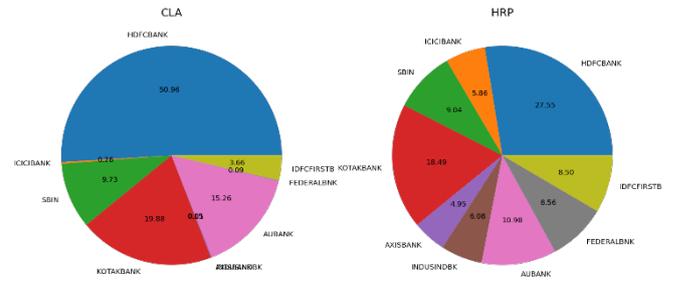

Fig. 6. Weight allocation to the auto sector stocks by the CLA and the HRP portfolio strategies (Training data: Jan 1, 2016 – Dec 31, 2020)

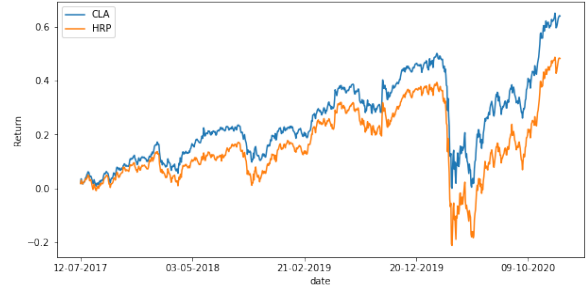

Fig. 7. Return yielded by the CLA and HRP portfolios on the training data for the banking sector stocks (Training data: Jan 1, 2016 – Dec 31, 2020)

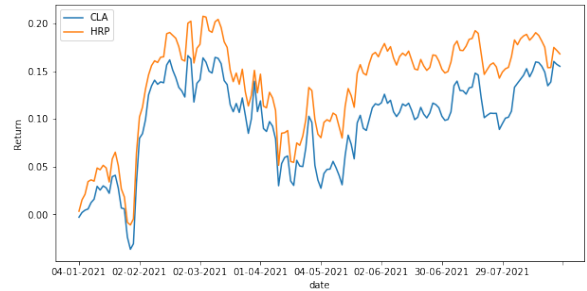

Fig. 8 Return from yielded by the CLA and HRP portfolios on the test data for the banking sector stocks (Test data: Jan 1, 2021 – Aug 26, 2021)

TABLE II. BANKING SECTOR PORTFOLIO PERFORMANCE

| Portfolio | Training Performance | | Test Performance | |
|---|---|---|---|---|
| | *Vol* | *Sharpe Ratio* | *Vol* | *Sharpe Ratio* |
| CLA | 0.2418 | 0.7789 | 0.2497 | 0.9786 |
| HRP | 0.2606 | 0.5447 | 0.2473 | 1.0713 |

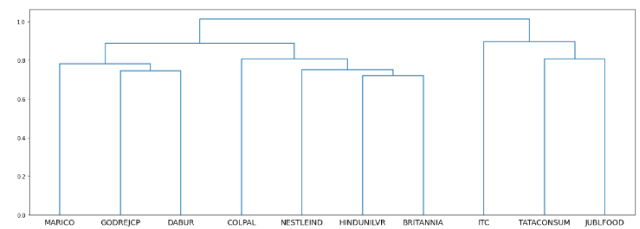

Fig. 9. The dendrogram of the agglomerative clustering of the FMCG sector stocks (Training data: Jan 1, 2016 – Dec 31, 2020)

*FMCG sector:* The ten critical stocks are as follows: Hindustan Unilever: 27.69, ITC: 23.80, Nestle India: 8.39, Tata Consumer Products: 6.02, Britannia Industries: 5.37, Godrej Consumer Products: 4.97, Dabur India: 4.66, Jubilant Foodworks: 3.84, Marico: 3.75, and Colgate Palmolive: 3.02 [2]. Fig 9 depicts the result of the agglomerative tree cluster on the FMCG sector stocks. There are three intuitive clusters formed. While Fig. 10 illustrates how the weights are allocated for the two methods, their performances on the in-sample and the out-of-sample data are depicted in Fig. 11 and Fig. 12, respectively. From Table III it is observed that the HRP has outperformed the CLA method on both training and test data yielding higher SR values for both cases.

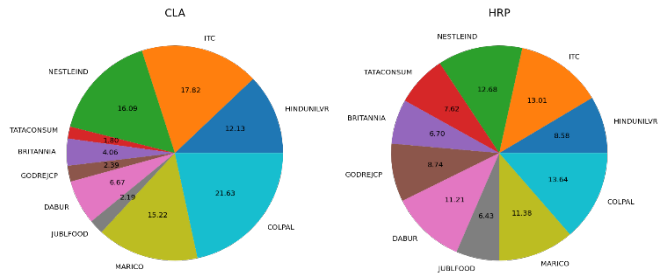

Fig. 10. Weight allocation to the FMCG sector stocks by the CLA and the HRP portfolio strategies (Training data: Jan 1, 2016 – Dec 31, 2020)

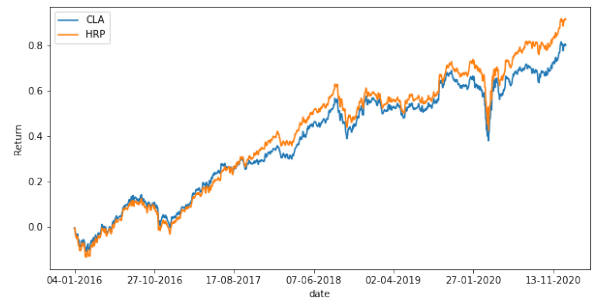

Fig. 11. Return yielded by the CLA and HRP portfolios on the training data for the FMCG sector stocks (Training data: Jan 1, 2016 – Dec 31, 2020)

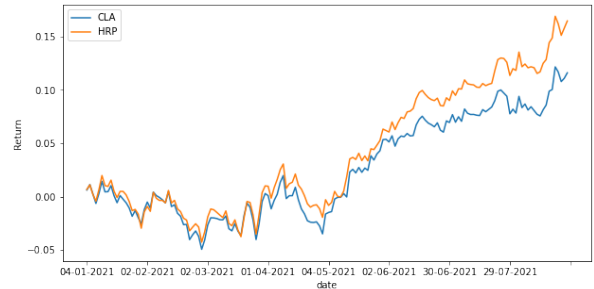

Fig. 12 Return from yielded by the CLA and HRP portfolios on the test data for the FMCG sector stocks (Test data: Jan 1, 2021 – Aug 26, 2021)

TABLE III. FMCG SECTOR PORTFOLIO PERFORMANCE

| Portfolio | Training Performance | | Test Performance | |
|---|---|---|---|---|
| | *Vol* | *Sharpe Ratio* | *Vol* | *Sharpe Ratio* |
| CLA | 0.1642 | 0.9941 | 0.1253 | 1.4569 |
| HRP | 0.1694 | 1.1003 | 0.1272 | 2.0365 |

*IT sector:* The critical stocks are: Infosys: 25.97, Tata Consultancy Services: 23.78, Tech Mahindra: 9.61, Wipro: 9.29, HCL Technologies: 8.91, MphasiS: 5.86, Larsen & Toubro Infotech: 5.83, MindTree: 5.04, Coforge: 2.95, Oracle Financial Services Software: 2.75 [2]. Fig 13 shows the results of the tree clustering on the IT sector stocks. There are three clusters formed, although the number of clusters finally will be determined by the threshold value of the linkage along the y-axis. The allocation of the weights as depicted in Fig 14, clearly shows that the HRP method has achieved much higher diversification in the portfolio. The performance of the two portfolios in terms of daily returns yielded by them on the in-sample and the out-of-sample data are shown in Fig. 15 and Fig 16, respectively. As observed from Table IV, The HRP portfolio has yielded higher values of SR on both training and test data than the corresponding values achieved by the CLA method.

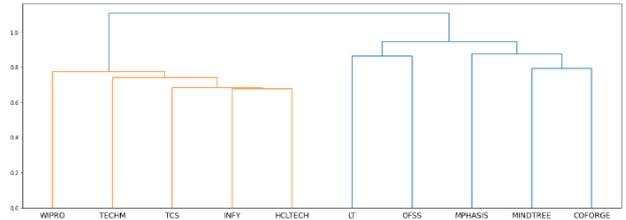

Fig. 13. The dendrogram of the agglomerative clustering of the IT sector stocks (Training data: Jan 1, 2016 – Dec 31, 2020)

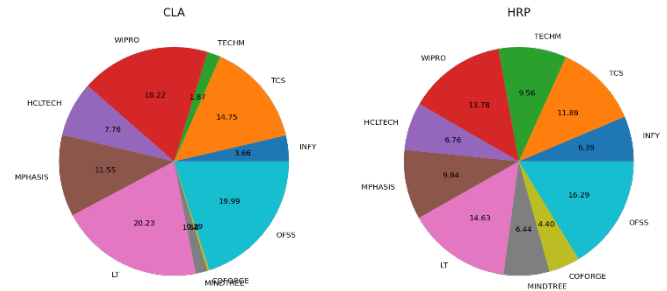

Fig. 14. Weight allocation to the IT sector stocks by the CLA and the HRP portfolio strategies (Training data: Jan 1, 2016 – Dec 31, 2020)

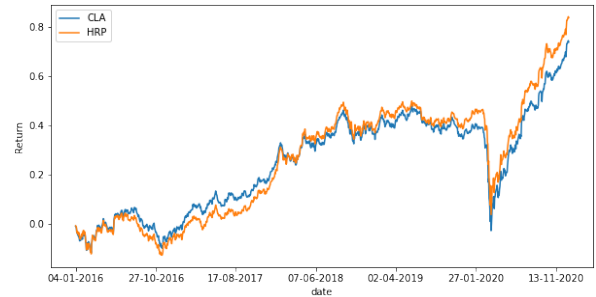

Fig. 15. Return yielded by the CLA and HRP portfolios on the training data for the IT sector stocks (Training data: Jan 1, 2016 – Dec 31, 2020)

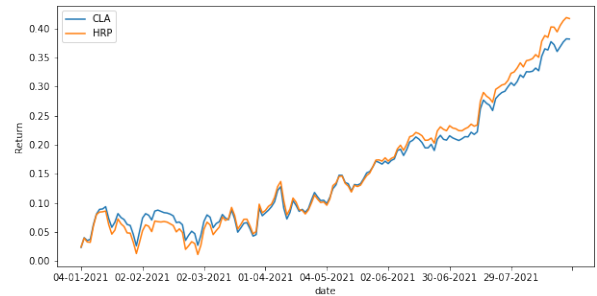

Fig. 16. Return yielded by the CLA and HRP portfolios on the test data for the IT sector stocks (Training data: Jan 1, 2016 – Dec 31, 2020)

TABLE IV. IT SECTOR PORTFOLIO PERFORMANCE

| Portfolio | Training Performance | | Test Performance | |
|---|---|---|---|---|
| | *Vol* | *Sharpe Ratio* | *Vol* | *Sharpe Ratio* |
| CLA | 0.1740 | 0.8668 | 0.1864 | 3.2244 |
| HRP | 0.1807 | 0.9466 | 0.1911 | 3.4354 |

*Metal sector:* This sector has the following ten most important stocks: Tata Steel: 24.9, JSW Steel: 15.85, Hindalco Industries: 14.45, Adani Enterprises: 8.69, Vedanta: 8.49, Coal India: 6.68, NMDC: 4.61, Steel Authority of India: 4.57, Jindal Steel & Power: 3.92, and APL Apollo Tubes: 2.52 [2]. Fig. 17 shows that the agglomerative clustering method yielding four intuitive clusters on the metal sector stocks. Fig 18 depicts that while the allocation of weights by the MPV method is highly skewed, a much higher diversification is achieved by the HRP weight allocation strategy. The benefits of this higher level of diversification are reflected in the superior performance of the HRP method on both training and test data; the SR values for the HRP method are higher than those of the CLA method in both cases.

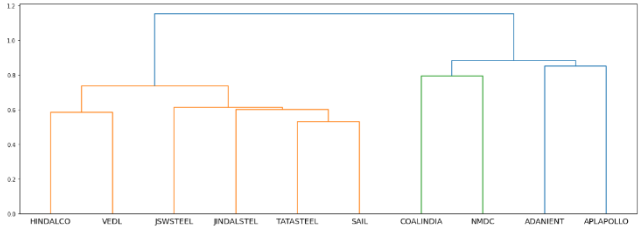

Fig. 17. The dendrogram of the agglomerative clustering of the metal sector stocks (Training data: Jan 1, 2016 – Dec 31, 2020)

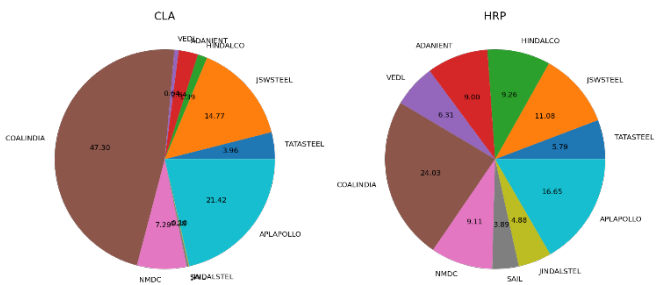

Fig. 18. Weight allocation to the metal sector stocks by the CLA and the HRP portfolio strategies (Training data: Jan 1, 2016 – Dec 31, 2020)

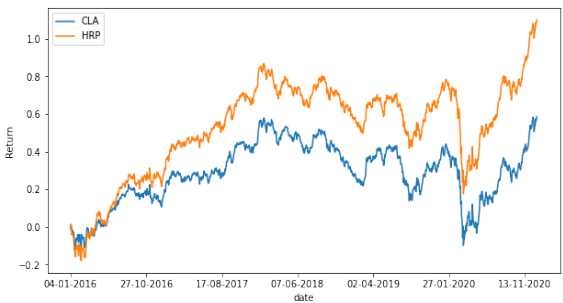

Fig. 19. Return yielded by the CLA and HRP portfolios on the training data for the metal sector stocks (Training data: Jan 1, 2016 – Dec 31, 2020)

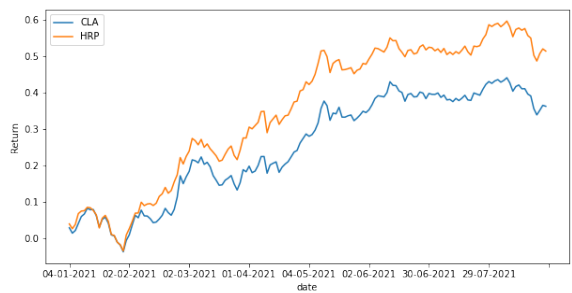

Fig. 20. Return yielded by the CLA and HRP portfolios on the test data for the metal sector stocks (Training data: Jan 1, 2016 – Dec 31, 2020)

*Pharmaceutical sector:* The important stocks of this sector are: Sun Pharmaceutical Industries: 22.87, Divi's Laboratories: 17.11, Dr. Reddy's Laboratories: 15.66, Cipla: 12.79, Lupin: 7.29, Aurobindo Pharma: 7.05, Biocon: 4.82, Alkem Laboratories: 4.18, Torrent Pharmaceuticals: 4.12, and Cadila Healthcare: 4.10 [2]. From Fig 21, it is observed that the agglomerative clustering algorithm has created three clusters among the pharma sector stocks based on their daily return values in the training data. The weight allocation done by the HRP method has created a more diversified portfolio as compared to the CLA method as evident from Fig. 22. The daily return yielded by the two portfolio approaches on the in-sample and the out-of-sample data are presented in Fig.23 and Fig. 24, respectively. However, the results tabulated in Table VI indicate that the performance of the CLA method is superior to that of the HRP method on both training and test data. The SR values for the CLA method are higher than those of the HRP method for both cases.

TABLE V. METAL SECTOR PORTFOLIO PERFORMANCE

| Portfolio | Training Performance | | Test Performance | |
|---|---|---|---|---|
| | *Vol* | *Sharpe Ratio* | *Vol* | *Sharpe Ratio* |
| CLA | 0.2432 | 0.4917 | 0.2592 | 2.2038 |
| HRP | 0.2698 | 0.8340 | 0.2790 | 2.9052 |

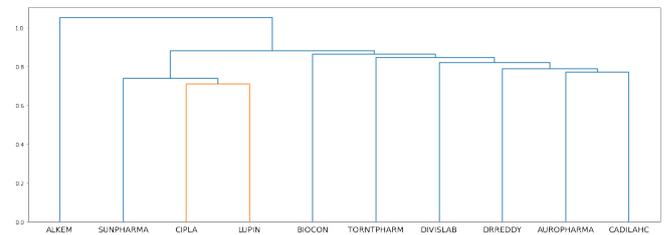

Fig. 21. The dendrogram of the agglomerative clustering of the pharmaceutical sector stocks (Training data: Jan 1, 2016 – Dec 31, 2020)

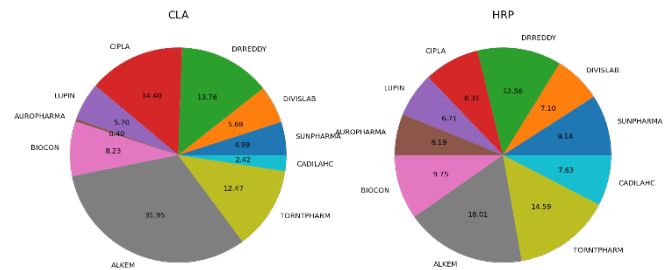

Fig. 22. Weight allocation to the pharmaceutical sector stocks by the CLA and the HRP portfolio strategies (Training data: Jan 1, 2016 – Dec 31, 2020)

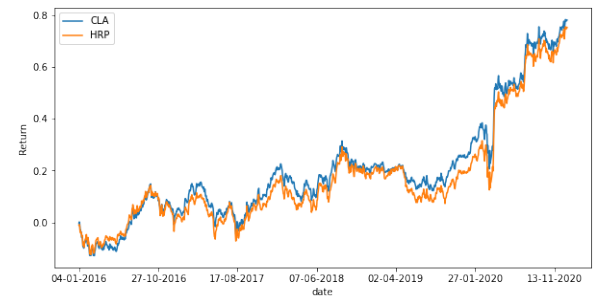

Fig. 23. Return yielded by the CLA and HRP portfolios on the training data for the pharmaceutical sector stocks (Trng data: Jan 1, 2016 – Dec 31, 2020)

*Realty sector:* The critical stocks of this sector are: DLF: 26.91, Godrej Properties: 24.09, Oberoi Realty: 10.42, Phoenix Mills: 10.32, Indiabulls Real Estate: 7.38, Prestige Estates Projects: 7.12, Brigade Enterprises: 6.25, Sobha: 3.20, Sunteck Realty: 2.74, and Hemisphere Properties India: 1.57 [2]. The agglomerative clustering algorithm has created eight clusters among the ten stocks of the realty sector as evident from Fig 25, although, the number of clusters will change based on the threshold value of distance chosen along the y-axis. The weight allocation done by the HRP achieved a higher diversification in the portfolio as evident from Fig. 26. The daily return values yielded by the two portfolios on the in-sample and the out-of-sample data are depicted in Fig. 27 and Fig. 28, respectively. It is clear from Table VII that while the HRP method has been outperformed by the CLA method on the training data because of the higher SR value yielded by the latter, the performance of the HRP method is superior to the CLA method on the test data, the SR value of the HRP on the test data is higher.

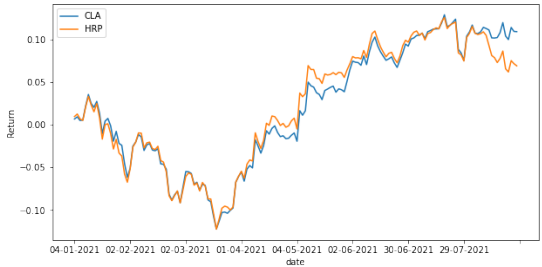

Fig. 24. Return yielded by the CLA and HRP portfolios on the test data for the pharmaceutical sector stocks (Trng data: Jan 1, 2016 – Dec 31, 2020)

TABLE VI. PHARMACEUTICAL SECTOR PORTFOLIO PERFORMANCE

| Portfolio | Training Performance | | Test Performance | |
|---|---|---|---|---|
| | *Vol* | *Sharpe Ratio* | *Vol* | *Sharpe Ratio* |
| CLA | 0.1839 | 0.8677 | 0.1739 | 0.9885 |
| HRP | 0.1939 | 0.7937 | 0.1794 | 0.6058 |

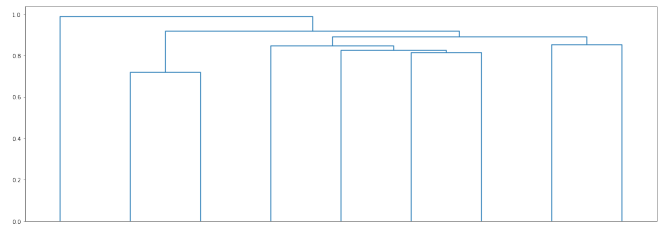

Fig. 25. The dendrogram of the agglomerative clustering of the realty sector stocks (Training data: Jan 1, 2016 – Dec 31, 2020)

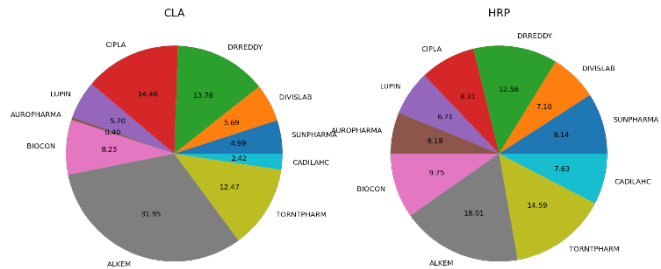

Fig. 26. Weight allocation to the realty sector stocks by the CLA and the HRP portfolio strategies (Training data: Jan 1, 2016 – Dec 31, 2020)

**NIFTY 50 stocks:** Finally, we consider the NIFTY 50 stocks and construct the CLA and the HRP portfolios based on the stocks. These stocks are the market leaders across 13 sectors and have a low-risk quotient. These 50 stocks are: "Asian Paints, Adani Ports, Axis Bank, Bharti Airtel, Bajaj Auto, Bharat Petroleum Corporation, Bajaj Finance, Britannia Industries, Coal India, Cipla, Divi's Laboratory, Eicher Motors, Dr. Reddy's Laboratories, Grasim Industries, HDFC, HCL Technologies, HDFC Life Insurance, Hindalco Industries, HDFC Bank, Hindustan Unilever, Hero MotoCorp, ICICI Bank, Infosys, IndusInd Bank, Indian Oil Corporation, JSW Steel, ITC, Kotak Bank, Mahindra & Mahindra, Larsen & Toubro, Maruti Suzuki, National Thermal Power Corporation, Nestle India, Oil and Natural Gas Corporation, Reliance Industries, Power Grid Corporation, State Bank of India, SBI Life Insurance, Shree Cement, Sun Pharmaceuticals, Tata Motors, Tata Steel, Tata Consultancy Services, Tech Mahindra, Tata Consumer Products, UltraTech Cement, Titan Company, Wipro, United Phosphorus[2]. The sectoral representation in percentage in the computation of the NIFTY 50 index is as follows. Financial Services: 37.58, IT: 18.01, Oil & Gas: 11.39, Consumer Goods: 11.05, Auto: 4.58, Metal: 3.75, Pharma: 3.46, Construction: 2.72, Cement & Cement Products: 2.50, Telecom: 2.11, Power: 1.55, Services: 0.74, and Fertilizers & Pesticides: 0.55 [2].

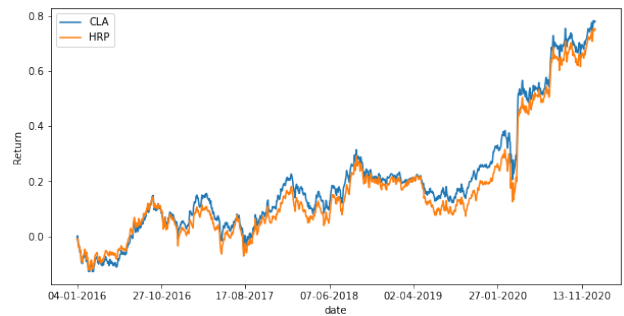

Fig. 27. Return yielded by the CLA and HRP portfolios on the training data for the realty sector stocks (Trng data: Jan 1, 2016 – Dec 31, 2020)

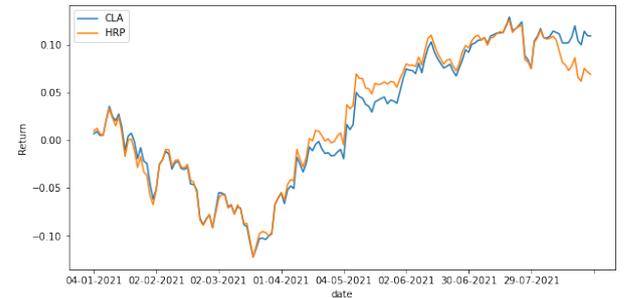

Fig. 28. Return yielded by the CLA and HRP portfolios on the test data for the realty sector stocks (Trng data: Jan 1, 2016 – Dec 31, 2020)

TABLE VII. REALTY SECTOR PORTFOLIO PERFORMANCE

| Portfolio | Training Performance | | Test Performance | |
|---|---|---|---|---|
| | *Vol* | *Sharpe Ratio* | *Vol* | *Sharpe Ratio* |
| CLA | 0.2493 | 1.0380 | 0.2670 | 1.1767 |
| HRP | 0.2548 | 0.9987 | 0.2671 | 1.3097 |

It is evident from Fig. 29 that there are three broad clusters created by the tree clustering algorithm on the NIFTY 50 stock data. Fig. 30 clearly shows that while the weights allocated by the CLA method are highly skewed (i.e., unequal), those assigned by the HRP are equal, resulting in a higher level of diversification in the portfolio. The daily return values of the two portfolios for the training and the test data are depicted in Fig. 31 and Fig. 32, respectively. The portfolio created from the NIFTY 50 stocks is expected to be a diverse one since there are candidate stocks for selection. With a higher level of diversification, the likelihood of a larger SR value for the HRP method is quite high. The results presented in Table VIII show that this is precisely the case. While the SR value for the HRP is lower than that of the CLA method on the training data, the reverse is the case for the test data.

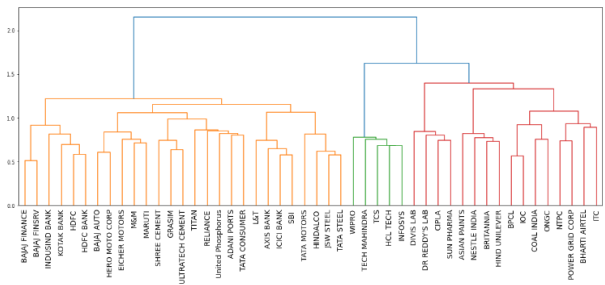

Fig. 29. The dendrogram of the agglomerative clustering of the NIFTY 50 stocks (Training data: Jan 1, 2016 – Dec 31, 2020)

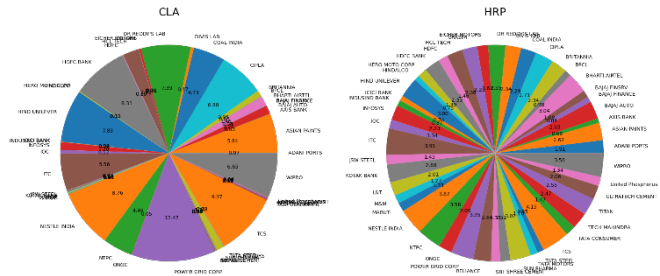

Fig. 30. Weight allocation to the NIFTY 50 stocks by the CLA and the HRP portfolio strategies (Training data: Jan 1, 2016 – Dec 31, 2020)

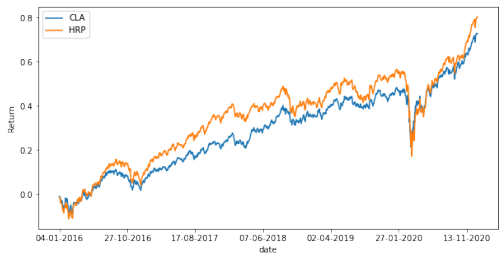

Fig. 31. Return yielded by the CLA and HRP portfolios on the training data for the NIFTY 50 sector stocks (Trng data: Jan 1, 2016 – Dec 31, 2020)

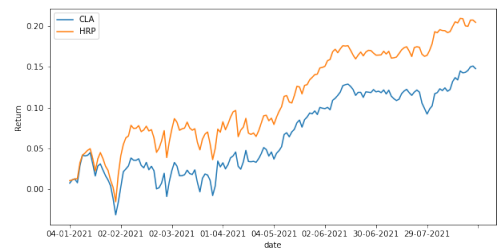

Fig. 32. Return yielded by the CLA and HRP portfolios on the test data for the NIFTY 50 stocks (Trng data: Jan 1, 2016 – Dec 31, 2020)

TABLE VII. NIFTY 50 PORTFOLIO PERFORMANCE

| Portfolio | Training Performance | | Test Performance | |
|---|---|---|---|---|
| | *Vol* | *Sharpe Ratio* | *Vol* | *Sharpe Ratio* |
| CLA | 0.1409 | 1.0552 | 0.1345 | 1.7292 |
| HRP | 0.1655 | 0.9913 | 0.1449 | 2.2213 |

*Summary results:* Table IX depicts a summary of the results, in which the portfolio yielding the higher SR value for a sector is mentioned for the training and the test case. While the CLA method has produced larger values of SR on the training data for 5 sectors out of 8 sectors studied in this work, the HRP method produced higher SR values for 7 sectors on the test data. The results clearly show that the performance of the HRP method is consistently better than that of the CLA method. This validates the theory of hierarchical risk parity allocation proposed in [2]. It may also be noted that on the out-sample data, SR is the maximum for the IT sector, while it is the minimum for the auto sector. Hence, for the period Jan 1, 2021, to Aug 26, 2021, IT sector stocks seemed to be the most profitable one while the auto sector seemed to have the lowest return.

TABLE IX. NIFTY 50 PORTFOLIO PERFORMANCE

| Sector | Training Performance | Test Performance |
|---|---|---|
| | *Portfolio with higher SR* | *Portfolio with higher SR* |
| Auto | CLA | HRP |
| Banking | CLA | HRP |
| FMCG | HRP | HRP |
| IT | HRP | HRP |
| Metal | HRP | HRP |
| Pharma | CLA | CLA |
| Realty | CLA | HRP |
| NIFTY 50 | CLA | HRP |

## V. CONCLUSION

This paper has presented portfolio design approaches on eight important Indian stock market sectors using the CLA and HRP algorithms. Exploiting the past prices of the top ten stocks of seven selected sectors and the 50 stocks from NIFTY50 of the NSE of India, the portfolios are designed. While the stock price data from Jan 1, 2016, to Dec 31, 2020, are used for building the portfolios, the period Jan 1, 2021, to Aug 26, 2021, is used for the testing. The portfolios are backtested on both the training and the test data, and the larger Sharpe Ratio-yielding portfolio is determined for each sector. It is found that while on the training data, the CLA approach yielded the higher Sharpe Ratios for five out of the eight sectors, the HRP method proved to be the clear choice on the test data delivering higher Sharpe Ratios for seven sectors.

REFERENCES

[1] H. Markowitz, "Portfolio selection", *Journal of Finance*, vol 7, no. 1, pp. 77-91, 1952.

[2] M. L. de Prado, "Building diversified portfolios that outperform out of sample", *Journ. of Portfolio Mgmt*, vol. 42, no. 4, pp. 59-69, 2016.

[3] E. Jurczenko, *Risk-Based and Factor Investing*, 1st edition, ISBN: 9781785480089, ISTE Press- Elsevier, 2015.

[4] D. Baily and M. L. de Prado, "Balanced baskets: A new approach to trading and hedging risks", *Journal of Investment Strategies.*, vol. 1, no. 4, pp. 61-62, 2012.

[5] NSE Website: http://www1.nseindia.com

[6] S. Mehtab, J. Sen, and A. Dutta, "Stock price prediction using machine learning and LSTM-based deep learning models", *Proc. of SoMMA*, pp. 88-106, 2020.

[7] S. Mehtab and J. Sen, "Stock price prediction using convolutional neural networks on a multivariate time series", *Proc. of NCMLAI*, Feb 2020.

[8] S. Metab and J. Sen, "A robust predictive model for stock price prediction using deep learning and natural language processing", *Proc. of 7th BAICONF*, Dec 5 -7, 2019.

[9] K. Liu, J. Zhou, and D. Dong, "Improving stock price prediction using the long short-term memory model combined with online social networks", *J of Behv. and Exp. Finance*, vol. 30, Jun 2021.

[10] Z. Wang, X. Zhang, Z. Zhang, and D. Sheng, "Credit portfolio optimization: A multiobjective genetic algorithm approach", Bora Istanbul Review, Jan 2021. (In press)

[11] J. Sen and S. Mehtab, "A comparative study of optimum risk portfolio and eigen portfolio on the Indian stock market", *Int. J. of Bus. Forcstng and Mktg. Intel.*, Inderscience Publishers, 2021 (In press).

[12] M. Corazza, G. di Tollo, G. Fasano, and R. Pesenti, "A novel hybrid PSO-based metaheuristic for costly portfolio selection problems". *Annals of Operations Research*, vol. 304, pp. 109-137, 2021.

[13] J. Sen, S. Mehtab, and A. Dutta, "Volatility modeling of stocks from selected sectors of the Indian economy using GARCH", *Proc. of the ASIANCON*, Aug 28-29, 2021 (To be listed in IEEE Xplore)

[14] J. Sen, S. Mehtab, and A. Dutta, "Stock portfolio optimization using a deep learning LSTM model", *Proc. of IEEE Mysurucon*, Oct 24-25, Mysore, India. (Accpeted for publication;to be listed in IEEE Xplore)